# Automatic Picking of P and S Arrivals Using a Minimum Uncertainty Wavelet Approach

Maria A. Krasnova

mkrasnova@ifz.ru

Institute of Physics of the Earth Russian Academy of Sciences

Bolshaya Gruzinskaya 10, Moscow, 123242, Russia

Donald J. Kouri

kouri@central.uh.edu

Department of Physics

University of Houston

Science and Research Bldg 1 3507

Cullen Blvd, Houston, TX 77204-5005

Evgeny Chesnokov

emchesno@central.uh.edu

Department of Earth and Atmospheric Sciences

University of Houston

Science and Research Bldg 1 3507

Cullen Blvd, Houston, TX 77204-5007

**Abstract**

The so-called "minimum uncertainty" or µ-wavelet approach has been shown to be very effective. Prior work has shown that using special algorithms based on a µ-wavelet decomposition of a seismic signal allows for arrival detection of various waves on a single-component seismogram with good precision. However, the question of interpretation of the type of wave arriving has not been addressed. The problem lies in the fact that the algorithm detects more than two arrivals even on seismograms with little noise, the peaks corresponding to P and S wave arrivals on the indicator function are not always the strongest, and the peak corresponding to the P wave arrival is not always the first peak detected. We propose to use a signal/noise relationship function in a running window as a weight coefficient of the indicator function obtained using the µ-wavelet transformation, which allows to isolate specifically those arrivals which correspond to significant changes in the signal amplitude. To interpret wave arrivals on linear registration systems (well observation systems for monitoring hydrolic fracturing, ultrasound investigations on models and samples of geological material), an algorithm was also developed to isolate arrivals that best correspond to arrivals on all registration points that have observation systems. As a result, we propose an algorithm that allows to automatically isolate arrivals of S and P waves on "frac" event recordings of a well observation system, and provides agreement with manual detection that doesn't exceed four discretes.

**Introduction**

*Arrival Indicators*

The volume of seismic information is currently so large that it renders manual processing of seismic recordings impractical. As a result, automated processing as well as detection and interpretation of seismic wave arrivals has become a necessity. In order to develop a reliable, automated algorithm to detect arrivals, it is necessary to describe the characteristic signs of wave arrivals in the seismic recording. In the case of a controlled source where the initial waveform of the seismic signal is known, the arrival detection reduces to the search for a particular waveform in the seismic recording. In the case of earthquake and microseismic data, the source is unknown. However, one can detect the following features of seismic recordings characteristic of P and S wave arrivals (see Figure 1):

1. A sudden increase occurs in the amplitude of the signal in the seismogram;
2. A sudden increase occurs in the frequency content of the signal in the seismogram.

Indeed, there is good reason to believe that the arrivals are "chirps", involving a rapid increase or decrease in the frequency content of the signal over a short time interval. In case of a recording from one separate, single-component station, one must rely solely on the above two characteristics for the detection of body wave arrival.

In the case of three-component recordings, the polarization of the body waves can be added to the signs of arrival. Two fundamental properties that are important for identifying arrivals are:

1. P and S wave arrivals are polarized orthogonally to each other (see Figure 2);

2. P wave's arrival is polarized linearly or quasi-linearly.

In processing the observation system recordings, we can also add indicators that are related to the observation system's geometry:

3. The difference in the timing of arrivals of the same wave at any two of the network stations cannot be longer than the travel time of the wave-type between the two stations.

The next indicators apply only to linear systems of observation:

4. For a vertical bore-hole system, the direction of the horizontal projection of the displacement in the first arrival of the P wave must be close on all stations;
5. The time graph of the wave's arrival along the profile must be adequately smooth (roughly, an approximation to a quadratic parabola) and not have strong oscillations.

All the above-enumerated indicators may be incorporated into an automatic scheme to detect arrivals.

*Using Wavelet Analysis to Detect Arrivals*

Since time and space are natural to us, seismic data is registered in the time domain, at specific locations in the earth, as $S_i(t), \ i = 1, 2, 3$ (three component detection) and is an oscillating function of time. A variety of processing methods exist that apply for different (e.g. frequency) recording domains. The main goal of such transformations is to find a domain in which the signal detection over the background of the seismic noise is easier than in time domain. The frequency domain, for example, can be used to perform a frequency filtration of the signal that is can improve the signal-to-noise (SNR) ratio. Since one of the signs of wave arrival is a very rapid change in the frequency composition of the signal (a chirp), it is expected that analysis of the signal in the combined time-frequency space will yield sharper signs of the wave arrival. However, this immediately raises the issue in Fourier theory that one cannot have arbitrary resolution in time and frequency simultaneously [Boashash, 1992]. This issue has been addressed by many and a very popular approach to optimally balancing simultaneous resolution in time and frequency is via "wavelets" (i.e., "little waves" when expressed in both the time and frequency domains [Mallat, 1999]). An approach that has been shown to be very effective is that of the so-called "minimum uncertainty" or µ-wavelets [Hoffman and Kouri, 2000].

Prior work by the Kouri group [Liao et al., 2011] showed that using special algorithms based on a µ-wavelet decomposition allows for arrival detection of various waves on a single-component seismogram with good precision. However, the question of interpretation of type of wave arriving was not addressed.

Schematically, the basic algorithm developed earlier [Liao et al., 2011] for detecting arrivals is presented below:

*A. Assume the signal at each station can be expressed as a linear combination of the **µ–wavelet***

$$S_{t_0}(t) = \sum_{j=0}^{N} C_j \varphi_j(t-t_0) \tag{1}$$

where we assume that the µ-wavelet, $\varphi_j(t-t_0)$, is localized in time at $t_0$. Thus, it only represents the signal in the neighborhood of the time $t_0$. We note that the wavelets are not orthogonal but they are linearly independent. In addition, they are also equally localized in the frequency domain.

*B. Transform the localized representation of the signal from time to frequency and compute the power of the localized signal*

The result is the detailed frequency content of the signal in the neighborhood of the time $t_0$, $S(t_0,\omega)$. We then compute the power spectrum received at the detector:

$$P(t_0,\omega) = \left| S(t_0,\omega) \right|^2, \tag{2}$$

and from this, we define the arrival detection function,

$$f(t_0) = \int d\omega \, P(t_0,\omega). \tag{3}$$

In the event that a seismic pulse arrives during this short time interval, we expect to see a large increase in the power. Sharp peaks in the arrival detection function suggest the arrivals of seismic energy.

It is clear that the analysis up to this point is not fully equipped to distinguish the types of waves that are arriving (apart from the general fact that the regular P-wave travels at a higher speed than do the S-waves). In addition, there is the question of false arrivals and the relative peak heights compared to noise. Thus, it is clear that the algorithm requires additional analysis to provide a clear picture of the arrival times of the body waves of interest.

In fact, the above is not a complete description of what must be done to actually apply the method. The algorithm is as follows:

Step 1. We cross-correlate (convolve) Eq. (1) with the µ-wavelet to obtain

$$\varphi_n(t) \otimes S(t) = \varphi_n(t) \otimes \sum_{j=0}^{N} C_j \varphi_j(t) \tag{4}$$

where

$$g(\tau) \otimes f(\tau) = \int_{-\infty}^{\infty} dt\, g(t+\tau) f(t) \tag{5}$$

We define

$$d_n = \int_{-\infty}^{\infty} dt\, \varphi_n(t+\tau) S(\tau) \tag{6}$$

and

$$X_{mn} = \int_{-\infty}^{\infty} dt\, \varphi_m(t+\tau) \varphi_n(t) \tag{7}$$

The equation for the contribution of the of the various wavelets can be expressed as

$$d_m(\tau) = \sum_n X_{mn}(\tau) C_n(\tau), \tag{8}$$

Then the coefficients $C_n$ can be obtained using

$$C_n(\tau) = \sum_l X_{ln}^+(\tau) d_l(\tau), \tag{9}$$

where $X_{ln}^+$ is the pseudo inverse of $X_{ln}$. (In general, $X_{ln}$ is ill-conditioned due to the non-orthogonality of the µ-wavelets.)

Step 2. We obtain the time-frequency representation of the signal by multiplying

the coefficients by Fourier transform of the $\mu$ wavelets

$$S(\tau, \omega) = \sum_n C_n(\tau)\, \varphi_n(\omega), \tag{10}$$

where

$$\varphi_n(\omega) = FT(\varphi_n(t)). \tag{11}$$

Step 3: After the power spectrum of the time-frequency form of the signal $P(\tau, \omega) = S^2(\tau, \omega)$ is defined, it is then integrated over frequency at various times $\tau$ to define the first break indicator function $f(\tau)$,

$$f(\tau) = \int S^2(\tau, \omega) d\omega. \tag{12}$$

Step 4: The next step is to apply a peak-picking algorithm to find the peaks, and denote them as an identifier for the first arrival. We specify a potential region of the first arrival. In this potential region, we neglect one other point and then apply the Hermite Distributed Approximating Functional (HDAFs) to fill in the neglected point [Hoffman and Kouri, 1995]. This effectively helps obtain a less noisy indicator.

*3. Wavelet family used in the analysis*

We here give the explicit form of the µ-wavelets. These functions are based on the standard Gaussian taking into account the minimization of the uncertainty principle $\Delta t\, \Delta \omega$ [Hoffman and Kouri, 2000]:

The µ - wavelet family is given by [Liao et al., 2011]:

$$\mu_j^\sigma(t) = \frac{1}{\sigma\sqrt{2j*j!\sqrt{\pi}}} H_j\left(\frac{\sqrt{\lambda}}{\sigma}t\right) e^{-\left(\frac{\sqrt{\lambda}}{\sigma}t\right)^2}, \tag{13}$$

where $H_j(x)$ is the j[th] order Hermite polynomial which is defined by

$$H_j(x) = (-1)^j e^{x^2} \frac{d^j}{dx^j} e^{-x^2}. \tag{14}$$

The λ and σ are parameters that control the bandwidth of the wavelets. We use the values $\lambda = 7$, $\sigma = 0.005$ [Liao, 2011].

*4. Amplitude-frequency characteristics of wavelets*

The amplitude-time and amplitude-frequency characteristics of the µ-wavelets are shown in Figure 3. To obtain spectral characteristics of wavelet functions, the standard Fourier transformation was used.

Earlier investigations [Liao et al., 2012; Kapur et al., 2014] showed that the µ-wavelet family allows for excellent precision in detecting wave arrivals, but the wavelets alone do not specifically identify P and S arrivals. The main goal of the present work is to modify the algorithm to process fully seismic recordings automatically (identifying P and P arrivals), implementing it in software. A software module was developed to determine arrivals in recordings of linear (borehole) three-component systems. Initially, it is necessary to choose two peaks which are interpreted as P and S wave arrival candidates from the multitudes of peaks in the indicator function. This same approach was used in the Liao study, where the two strongest peaks were selected for consideration. The analysis of the detected peaks in the indicator function the showed that:

1. The µ-wavelet analysis detects more than two wave arrivals, even on seismograms that have minimal noise levels and clear arrivals;
2. We believe that most, if not all of the detected peaks correspond to actual wave arrivals. These additional peaks may be arrivals of once-reflected or frequently-reflected converted waves from a single microseismic event, but also might be wave arrivals from different microseismic events;
3. Among all the arrivals, P and S wave arrivals are definitely detected.

The following conclusions can be made on the basis of detected arrival analysis:

1. Using the described µ-wavelet transformation algorithm allows for automatic wave arrival detection;
2. The algorithm detects more than two arrivals even on seismograms with little noise;
3. The peaks corresponding to P and S wave arrivals on the indicator function are not always the strongest, and the peak corresponding to the P wave arrival is not always the first peak detected.

Thus, the algorithm for choosing indicator function peaks corresponding to whole arrivals of body waves clearly requires modification and refinement. A modified algorithm based on combining the wavelet analysis with consideration of the signal-to-noise ratio is described below.

**Incorporating Signal-to-Noise Ratio Information**

It is important to remember that the µ-wavelet transformation allows us to detect chirp-like signals within a background of random seismic noise. However, in addition to the above indicator of body wave arrivals from microseismic events, drastic changes in the signal amplitude are also observed. Therefore, the peak search algorithm was augmented to consider the signal energies in two neighboring, running windows. The resulting additional indicator is called the signal-to-noise ratio function, $S2N_j(t_l)$. This quantity is defined as follows:

For each detector component ($j= 1,2,3$), the $S2N_j(t)$, $j=1$-$4$ functions are defined:

$$S2N_j(t_l) = \frac{\sum_{i=0}^{Nr} A_j^2(t_l+i \cdot dt)}{\sum_{i=0}^{Nl} A_j^2(t_l-i \cdot dt)}, for\ j = 1,2,3 \tag{15}$$

and

$$S2N_4(t_l) = \frac{\sum_{i=0}^{Nr}\left(\sum_j A_j^2(t_l+i \cdot dt)\right)}{\sum_{i=0}^{Nl}\left(\sum_j A_j^2(t_l-i \cdot dt)\right)}, \tag{16}$$

where $dt$ is the sampling frequency, $N_r = \frac{\Delta T_r}{dt}$, $\Delta T_r$ is the length of the right window (the signal window), $N_l = \frac{\Delta T_l}{dt}$, $\Delta T_l$ is the length of the left window (the noise window), and $A_j(t_l)$ is the amplitude of the $j$-th component at moment $t_l$. The function $S2Nj(t_l)$ is defined for times $T_l < t < T - T_r$.

The value of the S2N function is restricted as follows: if $S2N_j(t_l) < S2N_{min}$, then $S2N_j(t_l) = 0$. The length of the signal and noise windows as well as the minimum value of the function are adjustable parameters and may change for various types of data and observation ranges. The following were chosen for the current experiment:

- For microseismic recordings from a borehole observation system with sampling frequency of 4kHz, the length of the right window (the signal window) was chosen to be 5msec, and the length of the noise window was set at 7.5 msec;
- During the processing of seismic data obtained for a layered medium [Krasnova et al.,] (sharp arrivals, low noise level) with a sampling frequency of 2.5 MHz, the length of the signal and noise windows were chosen to be the same and equal to 4 μsec;
- $S2N_{min} = 1.6$ for both types of data.

Further, the "Modified Indicator Function", g(t), was introduced in terms of the indicator function $f(t)$ and the function $S2N(t)$:

$$g(t) = f(t) \cdot S2N^q(t). \tag{17}$$

In this definition, $S2N^q(t)$ can be regarded as a weight function accentuating peaks of the indicator function corresponding to the simultaneous appearance of two signs of wave arrival (appearance of a characteristic wave form and increase in the signal amplitude). Further, a standard algorithm of peak search and detection is applied to function *g(t)*. In this work values *q = 1, 2* were considered. For signals studied in this work, the best results were obtained for *q = 2*.

As before, the two maximal peaks of the indicator function were used as candidates for P and S arrivals. Comparison of the results comparisons for arrivals based on the indicator function *f(t)* and the generalized *g(t)* are shown on Figure 4 for laboratory results.

From the picture, it is evident that for an insignificant change in the array of peaks for each of the seismograms, the percent of automatically chosen arrivals corresponding to real ones rises significantly. That said, even on the seismograms where one or both max peaks do not correspond to real P and S arrivals, peaks corresponding to real arrivals are present.

**Adding a "Time Curve Smoothness" Criterion**

In addition to introducing the weight function *S2N(t)*, which allows us to improve the interpretation of arrivals at a single station, a module comparing arrival times obtained at separate stations and assessing the smoothness of the time curve was also added to the program. Presently, this module is implemented for a linear system of observations (the module was tested on an observation system in a vertical borehole), and the following assumptions are made for this function:

1. Peaks are identified using the indicator function *g(t)*, which correspond to P and S arrival times for least on 3 stations for P and at least on 3 stations for S. These sets of stations can be different for P and S waves;  At least at one station, one of two chosen arrivals must correspond to a real P wave arrival and at least on one station one must correspond to an S wave arrival.

2. The program module was set up for a linear system of observations with a constant distance between neighboring stations ($\Delta r = 40$ ft), placed at depths with an average velocity for P waves ~14000 ft/sec, and for S waves ~9000 ft/sec. (At present, these parameters can be changed only within the program code.)

The following arrival interpretation algorithm is used:

We consider an array of peaks detected on all the network's stations at times $t_i^k$, where $k = 1, 2, .., N_{stn}$, with $N_{stn}$ being the number of stations in a network, $i = 1, 2, .., N_{pks}^k$, with $N_{pks}^k$ being the number of peaks, detected at the $k$-th network station. For each of the stations, times of two maximal peaks $t_1^k$ and $t_2^k$ are chosen a priori as arrival times. Then the following procedure is applied stepwise to S and P waves:

Step 1. We find $t_{min} = \min_{k=1,..,N_{stn}} t_{1(2)}^k$, noting that this time corresponds to station $k_0$.

Step 2. Array $flag_k$ is entered, such that:

$$flag_k = \begin{cases} 1, & \text{if } t_{1(2)}^k - t_{min} \leq \frac{\Delta r \cdot |k-k_0|}{V_{P(S)}} \\ 0, & \text{if } t_{1(2)}^k - t_{min} > \frac{\Delta r \cdot |k-k_0|}{V_{P(S)}} \end{cases}. \qquad (18)$$

If more than three stations are found such that $flag_k = 1$, then we go to step 3.

If not, then on station $k$, $t_{1(2)}^k$ is assumed as the time of the next detected peak from this station from array $t_i^k$ and we return to step 1.

Step 3. On the chosen set $t_{1(2)}^k$ for which $flag_k = 1$, a quadratic parabola, which can be used as a rough approximation of P(S) wave time curve, is built using the least squares method: $t(k) = a_{1(2)}k^2 + b_{1(2)}k + c_{1(2)}$. $\qquad (19)$

Step 4. For stations for which $flag_k = 0$, a peak closest to the built parabola is chosen as an arrival for the array $t_i^k$. If no peak that is separate from the built time curve smaller than $\Delta t = \frac{\Delta r}{V_{cp}}$ exists in array $t_i^k$, then we assume that no wave arrival has been detected on this station.

As a control, during Step 3 we confirm that $a_{1(2)} > 0$ and $a_1 < a_2$ (i.e. that the velocity of the S wave is less than the velocity of the P wave).

The results of the time curve smoothness control module are shown on Figures 5 and 6.

Finally, Table 1 displays the arrival times detected automatically by the above-described algorithm, manually by the operator, as well as by the MSLocator program [Tomieka Searsy, et.al,2005].

As is evident from the figures and Table 1, the arrival time differences in arrivals detected manually and by the program described above does not exceed 3msec and most differences are within ±1 msec. This confirms that the automatic picking program and the manual operator basically chose the same arrival. Large differences with the MSLocator program show that this program detects and interprets different waves.

**Conclusions**

We have drawn the following conclusions from our study:

1. A successful program has been developed for automatic detection of P and S wave arrivals based on wavelet transformations for three wavelet families.
2. The wavelet program detects more than two wave arrivals even on seismograms with minimal noise and sharp arrivals. However, we conjecture that most, if not all detected peaks correspond to actual wave arrivals. These additional arrivals may correspond to once-reflected or frequently-reflected and converted waves from one microseismic, or perhaps wave arrivals from different microseismics.
3. Reliable P and S wave arrivals are detected, compared to a skilled manual picker.
4. The indicator function, obtained by a wavelet transformation and augmented by the weighted signal/noise function, allows one to focus on areas of the seismogram on which sharp changes in amplitude occur.

5. A procedure for time curve smoothness control is added which connects wave arrivals at all network stations.

6. Program results are compared against manual interpretation of arrivals, which shows that the program, in terms of arrivals, choses the same waves as the operator.

7. The mean difference in the automated arrival time detection and by the operator does not exceed 1msec (4 discretes), which is an acceptable precision in detecting wave arrivals.

**Acknowledgments**

Donald J. Kouri acknowledges partial support of this research under R. A. Welch Foundation Grant E-0608.

# Tables

Table #1. The arrival times detected automatically by the above-described algorithm, manually by the operator, as well as by the MSLocator program [Tomieka Searsy, et.al,2005].

| Event # | Station # | P wave arrivals | | | S wave arrivals | | |
|---|---|---|---|---|---|---|---|
| | | Manual | µ wavelet+ S2N+ time/distance | MSLocator | Manual | µ wavelet+ S2N+ time/distance | MSLocator |
| Frac event #295 (fig. 5) | 1 | 161.50 | 160.75 | 145.75 | 202.00 | 202.00 | 154.93 |
| | 2 | 160.75 | 161.25 | 144.29 | 200.00 | 200.00 | 154.48 |
| | 3 | 159.00 | 160.75 | 142.32 | 198.50 | 198.00 | 151.36 |
| | 4 | 159.00 | 159.00 | 142.86 | 197.75 | 198.00 | 148.00 |
| | 5 | 158.50 | 157.75 | 139.28 | 196.25 | 196.25 | 147.44 |
| | 6 | 158.25 | 158.50 | 136.17 | 195.00 | 194.75 | 141.26 |
| | 7 | 157.25 | 156.50 | 133.50 | 193.75 | 193.75 | 141.87 |
| | 8 | 158.00 | 157.75 | 135.50 | 194.25 | 195.25 | 139.01 |
| | 9 | 158.25 | 158.25 | 132.63 | 194.75 | 195.25 | 136.17 |
| | 10 | 158.50 | 158.25 | 132.00 | 195.00 | 194.00 | 132.25 |
| | 11 | 158.75 | 157.25 | 128.19 | 195.25 | 194.00 | 131.41 |
| | 12 | 159.50 | 159.75 | 126 | 195.50 | 195.50 | 126.25 |
| Frac event #210 (fig. 6) | 1 | 163.25 | 160.75 | 177.08 | 200.25 | 200.00 | 200.80 |
| | 2 | 159.25 | 158.25 | 173.75 | 197.25 | 197.00 | 196.74 |
| | 3 | 157.75 | --- | 175.50 | 198.75 | 195.25 | 195.79 |
| | 4 | 157.00 | 155.50 | 172.59 | --- | 195.75 | 195.00 |
| | 5 | 155.75 | 157.25 | 169.32 | 193.75 | 194.75 | 192.24 |
| | 6 | 155.75 | 155.25 | 170.50 | 193.25 | 193.25 | 190.32 |
| | 7 | 155.75 | 155.00 | 167.32 | 191.75 | 191.75 | 184.50 |
| | 8 | 156.25 | 155.75 | 164.90 | 192.00 | 192.00 | 183.75 |
| | 9 | 156.25 | 156.25 | 163.50 | 193.25 | 194.25 | 182.62 |
| | 10 | 156.75 | 154.50 | 160.50 | 191.75 | 192.25 | 179.75 |
| | 11 | 154.50 | 154.50 | 158.49 | 194.75 | 196.00 | 174.75 |
| | 12 | 156.50 | 156.50 | 158.33 | 193.75 | 193.75 | 173.84 |

**Figure captions**

Fig. 1 Seismic event record with detected P and S arrivals

Fig. 2 Polarization of P (blue) and S (green) waves.

Fig. 3 The amplitude-time and amplitude-frequency characteristics of 15 members of the μ-wavelet family.

Fig. 4 Comparing results of arrival detection for laboratory data on (a) the indicator function *f(t)* and (b) the generalized *g(t)*.

Fig. 5 Comparing results of arrival detection based on (a) the generalized function g(t) separately for each station and (b) using the time curve smoothness control module. Manual detection of P wave arrival – blue marks, S waves – green marks, automatically detected arrivals – red marks.

Fig. 6. One more example of arrival detection based (a) on the generalized function g(t) separately for each station and (b) using the time curve smoothness control module. Manual detection of P wave arrival – blue marks, S waves – green marks, automatically detected arrivals – red marks.

Figures will be supplied on request